# Selfconsistent Model of Photoconversion Efficiency for Multijunction Solar Cells

A.V. Sachenko[1], A.I. Shkrebtii[2], V.P. Kostylyov[1], M.R. Kulish[1], I.O. Sokolovskyi[1]

[1] V. Lashkaryov Institute of Semiconductor Physics, NAS of Ukraine, Kyiv, Ukraine
[2] University of Ontario Institute of Technology, Oshawa, ON, Canada

*Abstract* —To accurately calculate efficiencies $\eta$ of experimentally produced multijunction solar cells (MJSCs) and optimize their parameters, we offer semi-analytical photoconversion formalism that incorporates radiative recombination, Shockley-Read-Hall (SRH) recombination, surface recombination at the front and back surfaces of the cells, recombination in the space charge region (SCR) and the recombination at the heterojunction boundaries. Selfconsistent balance between the MJSC temperature and efficiency was imposed by jointly solving the equations for the photocurrent, photovoltage, and heat balance. Finally, we incorporate into the formalism the effect of additional photocurrent decrease with subcell number increase. It is shown that for an experimentally observed Shockley-Read-Hall lifetimes, the effect of re-absorption and re-emission of photons on MJSC efficiency can be neglected for non-concentrated radiation conditions. A significant efficiency $\eta$ increase can be achieved by improving the heat dissipation using radiators and bringing the MJSC emissivity to unity, that is closer to black body radiation rather than grey body radiation. Our calculated efficiencies compare well with other numerical results available and are consistent with the experimentally achieved efficiencies. The formalism can be used to optimize parameters of MJSCs for maximum photoconversion efficiency.

## I. Introduction

Much effort has recently been directed towards increasing efficiency $\eta$ of multijunction solar cells (MJSCs). MJSCs have shown promise for increasing photoconversion efficiency, therefore realistic optimization of MJSCs, given fundamental constraints on photoconversion, is important to provide achievable efficiency targets for both research labs and industry. However, standard theoretical formalisms of photoconversion are not comprehensive enough to realistically reproduce experimental efficiencies. Photoconversion efficiency $\eta$ of multijunction solar cells (MJSCs) under AM1.5 has reached 38.8% and 44.4% under concentrated illumination [1]. However, currently there is a gap between the experimental approaches to develop MJSCs, and often idealized formalisms to model their efficiency. The vast majority of existing computational work is performed based on thermodynamic Carnot approach or energy balance model (see, for example, [2] - [4]). Only one paper introduces such parameters as saturation currents for radiative recombination and Shockley-Read-Hall (SRH) recombination [5]. In the photoconversion model of [5] processes of light re-radiation and re-absorption (photon recycling) in the structures with multiple reflections were considered, resulting in open circuit voltage $V_{OC}$ increase. Recombination contribution to the photoconversion processes from radiative recombination and Shockley-Read-Hall recombination were also introduced in [5].

Meanwhile, to accurately and realistically predict efficiencies of experimentally produced MJSCs and optimize their parameters, photoconversion models must consider (i) radiative lifetime $\tau_r$, SRH lifetime $\tau_{SR}$, recombination in the space charge region (SCR) and surface recombination velocity $S$, (ii) optimal base and emitter doping levels, as well as their thickness, and (iii) selfconsistent balance between the MJSC temperature and efficiency. Since MJSCs contain multiple homo- or hetero-junctions, trade-offs between interfaces and thin film contributions are also important in the photoconversion processes, but they also were not sufficiently addressed.

To overcome the above mentioned limitations we develop a new semi-analytical selfconsistent approach to calculate photoconversion efficiency for MJSCs with both vertical and lateral design and apply the formalism to MJSC under AM0 and AM1.5 illuminations. In addition to radiative recombination [5], both Shockley-Read-Hall and interface recombinations (using typical values for direct bandgap III-V semiconductors) were included into the approach. We show that for the typical lifetime of order $10^{-8}$ s, which semiconductor technologies offer for $A_3B_5$ compounds based MJSCs, the contribution of the light re-absorption and re-emission to the photoconversion efficiency can be neglected.

In our simulation of the efficiency a balance between $\eta$ and the MJSC temperature $T$ is selfconsistently treated. With increase of subcell number $n$, the difference between the band gap and photon energies as well as the MJSC temperature decrease. This effect increases open circuit voltage $V_{OC}$ and efficiency $\eta$ and should be particularly strong for solar cells in space applications. In the outer space the solar cell temperature is not limited by reasonably high temperature of the environment such as in terrestrial application. Furthermore, it is demonstrated in selfconsistent way that the MJSC temperature strongly depends on the properties of heat sink that removes excessive heat from the MJSC.

Our formalism accounts for decrease of the emitted heat and operating MJSC temperature when the number of current-matched subcells $n$ increases. Increase of $n$ narrows the spectral range for each subcell, additionally reducing the photocurrent due to energy dependent light absorption, and this fundamental limitation is ignored in other approaches. This fun-damental effect reduces the limiting value of the photocurrent as the number of subcells $n$ increases. The photocurrent quantum yield $q_s(E_{ph})$ as a function of photon energy $E_{ph}$ near the absorption edge is below its maximal value of one. Therefore, if the solar spectrum is distributed over a larger number of subcells, the area below the curve $q_s(E_{ph})$, which is equal to the photocurrent in the photon energy range $\Delta E_{ph} = E_{ph2} - E_{ph1}$ for each subcell, declines more rapidly than the photon flux. Subsequently, despite the photovoltage increase with $n$ due to a larger number of energy intervals, the photocurrent starts decreasing due to lower quantum yield, thus causing the efficiency $\eta$ to saturate with the number of subcells. Existing theoretical approaches (see, *e.g.*, [5]), however, demonstrate only gradual increase of the efficiency $\eta(n)$. Our theoretical calculations agree well with experimentally available efficiencies.

The calculations were carried out for AM0 and AM1.5 conditions for both lateral (with the solar spectrum splitting), and vertical MJSC. We show that the improved heat dissipation from MJSC, in particular, the blackening of the back surface of the solar cell, and the use of radiators can significantly increase efficiency of MJSCs. The results presented in this study can be used to optimize the parameters of MJSC.

## II. FORMULATION OF THE PROBLEM

To calculate the efficiency of MJSCs and optimize their parameters we combine photoconversion physics laws in solar cells, which are governed by such characteristics of semiconductor materials as radiative lifetime $\tau_r$, the Shockley-Read-Hall time $\tau_{SR}$, surface recombination velocity $S$, doping levels of base and emitter as well as their thickness.

Our analysis shows that for SRH lifetimes $\tau_{SR}$ of order of $10^{-8}$ s, typical for technologically produced MJSCs based on $A_{III}B_V$ compounds, contribution of processes of re-emission and re-absorption of light (photon recycling) to the photoconversion efficiency can be neglected.

The thermal conductivity of subcells' materials is usually high enough, so the same stationary temperature is established throughout the whole MJSC under operation. Using results of [6], we formulated and solved a general thermal balance equation that comprehensively accounts for the main mechanisms of heat generation and dissipation though the black body emission and the heat sink. The absolute temperature of MJSC $T$ can be found from the equation:

$$P_s(r - \eta(T)) = \beta K_T \sigma (T^4 - T_{\min}^4) + \delta \cdot (T - T_{\min}) \quad (1)$$

with $P_s = \int_{E_1}^{E_2} P(E_{ph}) dE_{ph}$, where $P(E_{ph})$ is specific power of solar radiation at a given photon energy $E_{ph}$, $E_1$ and $E_2$ are the lower and upper photon energy limits for the entire MJSC, $J_m(T)$ and $V_m(T)$ are the photocurrent density and photovoltage at maximum power output. The emissivity parameter $\beta$ is of the order of unity and this defines a degree to which the MJSC radiates, compared to the perfect black body emission, and extra contributions to the heat up depending on the vicinity of the MJSC to the satellite and its orientation. Parameter $\sigma$ is the Stefan-Boltzmann constant, $T = T_{min} + \Delta T$, with $T_{min}$ being the ambient temperature, and $\delta$ is convection coefficient. In our calculations we considered that MJSCs absorb solar radiation in the wavelength range of $0.3\mu m < \lambda < 2\mu m$ (which corresponds to $4.13 \div 0.62$ eV photon energy range).

Coefficient $r$ accounts for incomplete ($r < 1$) absorption of light in the MJSC, which does not lead to increase its temperature, and that the radiative recombination takes away a part of the solar radiation energy, thus reducing the temperature. Convection coefficient $\delta$ depends on number of parameters such as wind speed, humidity, air density as well as type of material on which the solar cell is mounted. Therefore, actual MJSC temperature cannot be calculated by introducing average $\delta$ parameter. At the same time, even for sufficiently large convection coefficients (achieved, for instance, for high wind speed) MJSC temperature is always above the ambient temperature. However, the higher efficiency $\eta$ the closer MJSC temperature $T$ to $T_{min}$ is. Under AM1.5 $T_{min}$ is the ambient temperature, while for AM0 and low orbit satellites $T_{min} = 173$K [7]. In the cases of low wind speed or no wind the convection coefficient $\delta \sim 10^{-3}$ W/cm$^2$·K, and the radiative cooling is dominating if concentrators are used at AM1.5.

The open circuit voltage $V_{OC}$ of a single subcell, influenced by SRH recombination, recombination in SCR and the total surface recombination with its velocity $S_s$, can be calculated from the equation for the photocurrent density $J_g$:

$$J_g = q\left(\frac{d}{\tau_b} + S_s\right)\frac{n_i^2}{n_0}\left[\exp\left(\frac{qV_{oc}}{kT}\right) - 1\right] +$$

$$+ q\frac{\kappa L_D b^{-0.5} n_i}{\tau_{SR}\sqrt{\ln\left(\frac{\tau_{SR}}{q\kappa}\frac{J_g}{L_D n_i \exp(\frac{qV_{OC}}{2kT})}\right)}}\left[\exp\left(\frac{qV_{oc}}{2kT}\right) - 1\right]. \quad (2)$$

Here $q$ is elementary charge, $d$ is subcell thickness, $S_s$ is the total surface recombination velocity at the illuminated and back surfaces, $n_0$ is equilibrium concentration of the majority carriers, $\tau_b = (\tau_r^{-1} + \tau_{SR}^{-1})^{-1}$ is the bulk recombination lifetime, $\tau_r = (An_0)^{-1}$ is radiative lifetime with $A$ being the radiative recombination parameter, $\tau_{SR}$ is SRH recombination time, $\kappa \approx 2$, $L_D = (\varepsilon_0 \varepsilon_s kT / 2q^2 n_0)^{1/2}$ is the Debye screening length, $\varepsilon_0$ is the dielectric constant of vacuum, $\varepsilon_s$ is the relative permittivity of semiconductor, $k$ is Boltzmann constant, $b$ is the ratio of the hole capture cross section to the cross section for electron capture by the deep recombination center, $n_i(T) = \sqrt{N_c N_v}(T/300K)^{3/2} \exp(-E_g/2kT)$ is intrinsic carrier concentration, $N_c$ and $N_v$ are effective density of states of the conduction and valence bands at $T = 300$ K, and $E_g$ is the semiconductor bandgap.

Equation (2) is valid for $L > d$, where the diffusion length $L = (D \cdot \tau)^{1/2}$ with $D$ and $\tau$ are the minority carriers diffusion co-efficient and their lifetime respectively. The first term on the right side of (2) is written in the usual form [8], while the second term describes the contribution of different recombination mechanisms. This part is modified (as compared to [9]) for the case of illumination.

Eq. (2) can be rewritten in the following form:

$$J_{gi} = J_{0di} \exp\left(\frac{qV_{oci}}{kT}\right) + J_{0ri} \exp\left(\frac{qV_{oci}}{2kT}\right) = J_{0i}^* \exp\left(\frac{qV_{oci}}{m_i kT}\right). \quad (3)$$

Here $J_{0di}$ is the diffusion saturation current density, $J_{0ri}$ is the recombination saturation current density, $J^*_{0i}$ the effective saturation current density, and $m_i$ is the effective ideality factor of the I-V curve for the subcell under consideration. Comparing (2) and (3) we can find $J_{0di}$ and $J_{0ri}$, modified by the above considered recombination mechanisms.

According to [10] the open circuit voltage $V_s$ of MJSC, that is for the series-connected subcells, can be written:

$$V_s = \frac{kT}{q}\sum_{i=1}^{n} \ln\left(\frac{J_g}{J_{0i}^*}\right)^{m_i} = m_s \frac{kT}{q} \ln\left(\frac{J_g}{J_{0s}}\right), \quad (4)$$

with

$$m_s = m_1 + m_2 + \ldots m_n, \quad (5)$$

$$J_{0s} = \left(J_{01}^{*m_1} \cdot J_{02}^{*m_2} \cdot \ldots \cdot J_{0n}^{*m_n}\right)^{\frac{1}{m_s}}. \quad (6)$$

The current density $J$ in the $I-V$ characteristics for the series connected subcells in this case depends on voltage $V$ as:

$$J(V) = J_g - J_{0s} \exp\left(\frac{qV}{m_s kT}\right). \quad (7)$$

Applied voltage $V_m$, which yields the maximum MJSC power output $P_m$ ($P_m = J_m \cdot V_m$), can be found from the following:

$$\frac{d}{dV}(J(V)V) = 0, \quad (8)$$

which gives us

$$V_m \cong V_s\left(1 - \frac{m_s kT \ln(qV_m/m_s kT)}{qV_m}\right). \quad (9)$$

Short-circuit current density can be found for both lateral and vertical MJSCs. For the lateral MJSCs the short-circuit current densities of the single $i$-th cell $J_{gi}^L$ can be written:

$$J_{gi}^L(E_{gi}) = s_i^{-1} \int_{E_1}^{E_2} j_{gi}(E_{ph}) q_{si}(E_{gi}, E_{ph}) dE_{ph}. \quad (10)$$

Here parameter $s_i = S_i/S$, is the specific surface area of the $i$-th subcell $j_{gi}(E_{ph})$ is the photocurrent density at photon energy $E_{ph}$, which is a product of elemental charge $q$ and solar radiation flux density $P(E_{ph})$, and $q_{si}(E_{gi}, E_{ph})$ is the photocurrent quantum yield.

The general expression for $q_{si}(E_{gi}, E_{ph})$ is given in [8], which can be rewritten for our case as:

$$q_s = q_{sp} + q_{sn}, \quad (12)$$

where

$$q_{sp} = \frac{\alpha L_p}{(\alpha L_p)^2 - 1} \cdot \frac{1}{S_0 \frac{\tau_p}{L_p}\sinh\left(\frac{d_p}{L_p}\right) + \cosh\left(\frac{d_p}{L_p}\right)} \cdot [\alpha L_p + S_0 \frac{\tau_p}{L_p}\left(1 - e^{-\alpha d_p}\right)\cosh\left(\frac{d_p}{L_p}\right) - e^{-\alpha d_p}\sinh\left(\frac{d_p}{L_p}\right) - \alpha L_p e^{-\alpha d_p}], \quad (13)$$

and

$$q_{sn} = \frac{\alpha L e^{-\alpha d_p}}{1-(\alpha L)^2} \cdot \frac{1}{S_d \sinh\left(\frac{d}{L}\right) + \frac{D}{L}\cosh\left(\frac{d}{L}\right)} \left\{[S_d \cosh\left(\frac{d}{L}\right) + \frac{D}{L}\sinh\left(\frac{d}{L}\right)](1+R_d e^{-2\alpha d}) + (\alpha D(1-R_d) - S_d(1+R_d))e^{-\alpha d} - \alpha L S_d\left[\sinh\left(\frac{d}{L}\right) + \frac{D}{L}\cosh\left(\frac{d}{L}\right)\right](1 - R_d e^{-2\alpha d})\right\}. \quad (14)$$

Here $\alpha$ is the light absorption coefficient, $L_p$ is the diffusion length in the emitter of width $d_p$ and bulk recombination time $\tau_p$, and $R_d$ stands for back surface reflection coefficient.

For the vertical MJSC the expression for the short-circuit current density $J_{gi}^V$ has the form:

$$J_{gi}^V(E_{gi}) = \int_{E_1}^{E_2} j_g(E_{ph}) q_{si}(E_{gi}, E_{ph}) T_i(E_{gi}, E_{ph}) dE_{ph}, \quad (15)$$

and, with $\alpha_i$ being the light absorption coefficient for $i$-th cell,

$$T_i(E_{gi}, E_{ph}) = e^{-\alpha_1 d_1 - \alpha_2 d_2 \cdots - \alpha_{i-1} d_{i-1}}. \quad (16)$$

For the quantum yield of $i$-th subcell the expressions (10) - (12) can be used by putting the reflection coefficient from $i$-th subcell back surface $R_{di}=0$.

For series-interconnected subcells photocurrent-matching condition has to be satisfied. The current matching implies that the photocurrent of each subcell is the same, *i.e.*:

$$J_{g1}(E_{g1}) = J_{g2}(E_{g2}) = \cdots = J_{gn}(E_{gn}) = J_g. \quad (17)$$

The photocurrent density at the point of maximum power output $J_m$ can be found as:

$$J_m = J_g(1 - m_s kT/qV_m). \quad (18)$$

Considering the above recombination losses and ignoring for simplicity light reflection and shading, as well as losses due to dispersion element, the corresponding efficiency $\eta_0$ is:

$$\eta_0 = J_m V_m / P_s. \quad (19)$$

Solving selfconsistently the set of equations together with expressions (1) – (19) the MJSC efficiency $\eta$ equals:

$$\eta = Q_{dr}(1 - R_s)(1 - K_m)\eta_0, \quad (20)$$

where $Q_{dr}$ is the dispersion element efficiency, $R_s$ is the coefficient of the light reflection from the MJSC front surface and $K_m$ is the shading coefficient. In the numerical solution for the simplicity, we consider that $R_s =0$, and $K_m=1$.

III. RESULTS AND ANALYSIS

Since the solar spectrum that reaches subcells of MJSC and being adsorbed is in the 0.3μm < $\lambda$ < 2μm wavelengths range, non-photoactive low energy photons with $\lambda$ > 2μm do not contribute to the SC temperature increase. For simplicity of our analysis, we consider here the recombination losses only, and in such a case it follows from Eg. (20) that $\eta(n) = \eta_0(n)$.

Processes of re-absorption of light are considered in [5] are important if the nonradiative recombination can be neglected. When calculating $V_{OC}$ in this case, instead of the radiative recombination parameter $A$ we have to use $A_{eff}=A(1-\gamma_r)$ with $\gamma_r$ the photon re-emission factor. This increases the efficiency, and when we calculated $\eta$, the reflection coefficient $R_d$ was put to zero. In contrast, when the nonradiative recombination is present, $\eta$ change due to the re-absorption can usually be neglected [8]. In this paper, rather than relying on re-absorption processes, the MJSC efficiency gain is achieved mainly by improving heat dissipation. This can be done by increasing the MJSC emissivity parameter $\beta$, making this as close to blackbody as possible. If the solar cell heat dissipates through the top illuminated surface, $\beta$ is close to 0.9 [11]. By blackening the back surface of the MJSC, the emissivity parameter can be increased to ~1.8. Another method of improving the heat dissipation is to use radiator fins, where the heat emission occurs from larger area than the area of the illuminated MJSC surface ($K_T$>1).

We also analyze the influence of the subcells' base doping level $n_0$ on MJSC photoconversion efficiency $\eta$. When the Shockley-Read-Hall recombination dominates, $\eta$ reaches maximum at $n_0 = 10^{17}$cm$^{-3}$. For the $n$-type base the maximum of $\eta$ is due to interplay of both Auger recombination and Burstein-Moss effect, with the last decreasing the light absorption coefficient near the absorption edge. For the $p$-type base, Burstein-Moss effect does not contribute to the maximum of efficiency happening. The above considerations are used to optimize $n_0$ when simulating MJSC efficiency $\eta$.

We first calculate $\eta(n)$ for the *hypothetical model system* when each subcell is made of a direct-gap III-V semiconductor with exact energy gaps $E_{gi}$, always available to satisfy current matching. The following parameters were used: $A_{eff}$ = $2 \cdot 10^{-10}$ cm$^3$/s, $D$ = 50 cm$^3$/s, $N_c$= $5 \cdot 10^{17}$ cm$^{-3}$, $N_v$= $10^{19}$ cm$^{-3}$. We also considered $d = 2 \cdot 10^{-4}$ cm, $d_p = 10^{-6}$ cm, $\tau_p = 10^{-10}$ s, $n_0 = 10^{17}$ cm$^{-3}$, $S_s = 10^3$ cm/s and $\delta = 10^{-2}$ W/sm$^2 \cdot$K. The surface recombination velocities and the Shockley-Read lifetimes considered to be the same for each subcells of MJSC. When SRH recombination is considered, we use $\tau_{SR} = 5 \cdot 10^{-9}$ s.

Figs. 1a and 1b demonstrate calculated efficiencies $\eta(n)$ for the lateral MJSC for a large number of subcells $n > 10$ under AM0 and AM1.5 respectively. Comparing Figs. 1a and 1b, the efficiency at AM0 is always higher than at AM1.5, which is due to lower MJSC temperature under AM0 condition. The better heat dissipation (which corresponds to high value of $\gamma=\beta K_T$) the lower MJSC temperature is and the higher is the efficiency, graphs 1 – 3. The graph 4 in Fig. 1b is taken from [5] and corresponds to recombination saturation current in SCR equals 100 mA/cm$^2$. It appears from Fig. 1b that the efficiency $\eta$ from [5] is close to our theoretical curve 2.

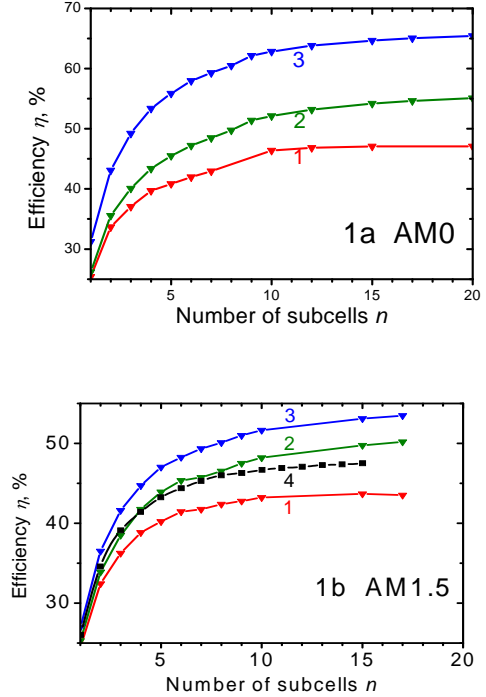

Fig. 1. (Color online). Efficiency $\eta(n)$ of lateral MJSC at AM0 (Fig. 1a) and AM1.5 (Fig. 1b) for hypothetical model system for different heat dissipation conditions: $K_T$=1 for all curves, $\beta$ values are 1, 1.5 and 2 for curves 1, 2 and 3 respectively.

In contrast to the previous figure, Fig. 2 shows $\eta(n)$ at AM1.5 for a specific set of *technologically available* semiconductors with predefined gaps $E_g$: GaSb, GaAs$_{0.7}$Sb$_{0.3}$, GaAs, Al$_{0.15}$Ga$_{0.85}$As, Al$_{0.3}$Ga$_{0.7}$As, and Al$_{0.45}$Ga$_{0.55}$As. Here, $R_d$=0 and $\beta$ =1.5, while the heat dissipation conditions are varied: curve 3 $K_T$=1, curve 2 $K_T$=5, and curve 3 corresponds to a forced cooling with MJSC temperature $T$=300K, which produces the highest $\eta$. Curve 4 describes the experimentally achieved efficiencies for different numbers of subcells in MJSC [12], [13] [14]. The figure demonstrates reasonably good agreement of our theoretical results with experimentally available efficiencies. Comparing Figs. 1 and 2 it is important to stress that instead of saturation of $\eta(n)$ dependences for model system (Fig. 1), the theoretical $\eta(n)$ graphs clearly show maximum when parameters of technologically available semiconductors are considered (Fig. 2).

It is important to stress that for improved heat sink, for instance, with $\gamma = 6 \cdot 10^{-3}$ W/cm$^2$·K [15] at AM1.5 MJSC temperature increases only slightly above the ambient temperature, thus resulting into to the efficiency close to that one at ambient temperature.

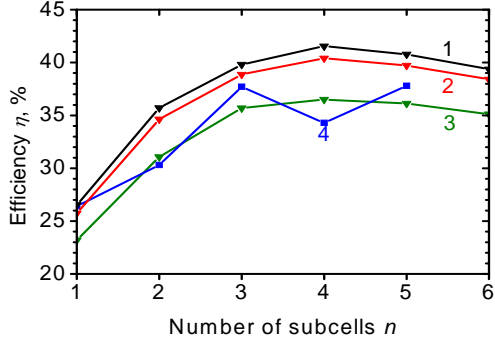

Fig. 2. (Color online). Calculated (curves 1 – 3) vs. experimental (curve 4) MJSC efficiencies $\eta(n)$ at AM1.5, $R_d$=0 and $\beta$ =1.5. For the curve 1 we neglect the surface recombination, while it is included for curves 2 and 3. Curve 4 corresponds to experimentally achieved efficiencies [13], [14], and [16] for the following set of semiconductors: GaSb, GaAs$_{0.7}$Sb$_{0.3}$, GaAs, Al$_{0.15}$Ga$_{0.85}$As, Al$_{0.3}$Ga$_{0.7}$As, and Al$_{0.45}$Ga$_{0.55}$As.

Three upper curves in Fig. 3 show calculated maximum attainable MJSC efficiencies $\eta$ as a function of subcell number $n$ for model system. Curve 1 is calculated within the energy balance model, taken from [11]. Curve 2 corresponds to the highest calculated efficiency in [5] (see Fig. 19, uppermost curve when the only recombination mechanism is the radiative recombination). Curve 3 is our calculated MJSC efficiency $\eta$ for the case of improved heat dissipation and all recombination mechanisms, except radiative recombination, neglected.

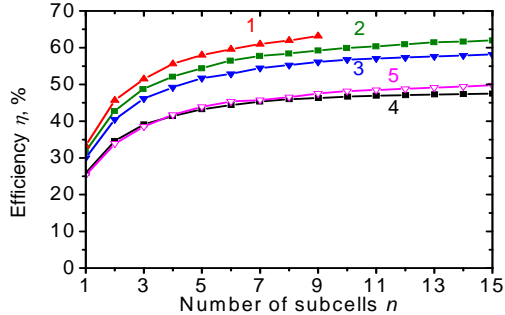

Fig. 3. (Color online). Calculated efficiencies for model system with set of semiconductors under AM1.5, curve 1 is from [5], curve 2 is from [11] and curve 3 is from our approach. Details are in the text.

Curves 1 – 3 in Fig. 3 correspond to the maximum attainable MJSC efficiency, but calculated using different formalisms. The fact that curves 1 – 3 are within 10% of each other, demonstrates that efficiencies $\eta$, calculated using these different formalisms, converged to close values.

Curves 4 and 5 in Fig. 3 describe the efficiency when Shockley-Read-Hall recombination is predominant. Curve 4 is from [5], Fig. 19 with SRH recombination current density of 100 A/cm$^2$. Curve 5 simply corresponds to curve 4 in Fig. 1b, calculated using our approach. We stress that the similarity of the curves 4 and 5 is important since they have been calculated using different formalisms. However, our transparent and simple to implement approach, based on general Eq. (2), is more efficient and universal.

We restricted here our analysis of $\eta(n)$ to non-concentrated solar radiation, although the formalism developed is applicable to the case of concentrated light as well. We only note that when using concentrators, to increase MJSC efficiencies, the degree of concentration $M$ has to be proportional to $K_T$. Otherwise, increasing $M$ will substantially heat up MJSC, which lead to $\eta$ decrease. Obviously, sufficiently high $K_T$ can be only achieved under AM1.5. In the outer space producing systems with high $K_T$ is difficult due to MJSC weight restrictions.

Unlike [5], in our formalism, the MJSC efficiency increase is achieved through improved heat dissipation rather than due to multiple reflections and re-absorption of the light. The approach of [5] can be beneficial when Shockley-Reed-Hall recombination is negligible compared to the radiative recombination. Our formalism is not only advantageous when SRH recombination dominates over radiative recombination, but also includes SCR and surface recombination.

IV. CONCLUSIONS

In this paper we propose modified efficient semi-analytical approach to realistically calculate the photoconversion efficiency $\eta(n)$ for multijunction solar cells, applicable for both terrestrial and space applications, and using concentrated or non-concentrated light. The formalism incorporates important mechanisms of photogenerated carriers' losses due to Shockley-Read-Hall bulk recombination, space charge layer recombination as well as surface recombination at the illuminated and back MJSC surfaces, and heterojunction interfaces respectively. The above recombination contributions reduce the efficiency $\eta$ compared to the case when only radiative recombination dominates. MJSC efficiency $\eta(n)$ has been calculated first for the hypothetical "ideal" system when each subcell is made of a direct-gap III-V semiconductor with exact energy gap $E_{gi}$, required to satisfy current matching. With efficient heat sink and increasing subcells number $n$, calculated maximum MJSC efficiencies can exceed 60% for AM0 conditions and 50% for AM1.5.

In contrast, applying our formalism to experimentally available semiconductors with predefined energy gaps $E_g$ for se-lected combinations of materials, the maximum efficiency achieved at AM1.5 $\eta(n)$ is close to 41% at $n = 4$. The calculated efficiencies agree reasonably well with the experimental results available in the literature. These findings demonstrate that there still exists sufficient ability to substantially increase MJSC efficiency by considering direct-gap semiconductors with best match of the bandgaps and number of subcells.

Calculated at AM1.5 efficiencies $\eta(n)$ depend not only on specific selection of MJSC subcells, but also on non-monotonic energy dependence for the solar spectrum.

Our calculated efficiencies agree reasonably well with those calculated within different formalisms. In contrast to [5], we conclude that MJSC efficiency $\eta$ increase is not due to multiple reflections and reabsorption of light, but due to improved heat management of the system. Results of [5] can be used in a case when SRH recombination is insignificant compared to the radiative recombination, which corresponds to very high level of injection, not achievable in the real solar cells. Advantages of our approach described in this paper are that we considered more realistic conditions when SRH recombination dominates over the radiative recombination. This corresponds to, for instance, AM1.5 conditions.